\documentclass[twocolumn]{aastex63} \usepackage{amsmath} \usepackage{booktabs}
\usepackage[utf8]{inputenc} \usepackage{xcolor} \usepackage{graphicx} \usepackage{hyperref}
\newcommand{\FIXME}[1]{\textcolor{black}{#1}}

\newcommand{\likelihood}{\ensuremath{\log \mathcal{L}}}
\newcommand{\SNRThresh}{\FIXME{4.0}}

\newcommand{\NumAcceptedSims}{\FIXME{$259\thinspace200$}}
\newcommand{\NumRejectedSims}{\FIXME{$1\thinspace659\thinspace747$}}
\newcommand{\TotalSims}{\FIXME{$1\thinspace918\thinspace947$}}

\newcommand{\SimMinMass}{\FIXME{$1.0 \, M_\odot$}}
\newcommand{\SimMaxMass}{\FIXME{$2.0 \, M_\odot$}}
\newcommand{\SimMeanMass}{\FIXME{$1.33 \, M_\odot$}}
\newcommand{\SimStdev}{\FIXME{$0.09 \, M_\odot$}}
\newcommand{\SimMaxRedshift}{\FIXME{0.2}}

\newcommand{\BankSize}{\FIXME{$80\thinspace679$}}
\newcommand{\BankMM}{\FIXME{98\%}}
\newcommand{\BankMinMass}{\FIXME{$0.95 M_\odot$}}
\newcommand{\BankMaxMass}{\FIXME{$2.4 M_\odot$}}
\newcommand{\BankMinChirp}{\FIXME{$0.9 M_\odot$}}
\newcommand{\BankMaxChirp}{\FIXME{$1.7 M_\odot$}}

\begin{document}

\title{An early warning system for electromagnetic follow-up of gravitational-wave events}

\author{Surabhi Sachdev} 
	\affiliation{Department of Physics, The Pennsylvania State University, University Park, PA 16802, USA}
	\affiliation{Institute for Gravitation and the Cosmos, The Pennsylvania State University, University Park, PA 16802, USA}

\author{Ryan Magee} 
	\affiliation{Department of Physics, The Pennsylvania State University, University Park, PA 16802, USA}
	\affiliation{Institute for Gravitation and the Cosmos, The Pennsylvania State University, University Park, PA 16802, USA}

\author{Chad Hanna} 
	\affiliation{Department of Physics, The Pennsylvania State University, University Park, PA 16802, USA}
	\affiliation{Institute for Gravitation and the Cosmos, The Pennsylvania State University, University Park, PA 16802, USA}
	\affiliation{Department of Astronomy and Astrophysics, The Pennsylvania State University, University Park, PA 16802, USA}
	\affiliation{Institute for CyberScience, The Pennsylvania State University, University Park, PA 16802, USA}

\author{Kipp Cannon} 
	\affiliation{The University of Tokyo, Hongo 7-3-1 Bunkyo-ku, Tokyo 113-0033, Japan}

\author{Leo Singer}
	\affiliation{NASA/Goddard Space Flight Center, Greenbelt, MD 20771, USA}

\author{Javed Rana SK} 
	\affiliation{Department of Physics, The Pennsylvania State University, University Park, PA 16802, USA}
        \affiliation{Institute for Gravitation and the Cosmos, The Pennsylvania State University, University Park, PA 16802, USA}

\author{Debnandini Mukherjee} 
	\affiliation{Department of Physics, The Pennsylvania State University, University Park, PA 16802, USA}
        \affiliation{Institute for Gravitation and the Cosmos, The Pennsylvania State University, University Park, PA 16802, USA}

\author{Sarah Caudill}
	\affiliation{Nikhef, Science Park, 1098 XG Amsterdam, Netherlands}	

\author{Chiwai Chan}
	\affiliation{The University of Tokyo, Hongo 7-3-1 Bunkyo-ku, Tokyo 113-0033, Japan}

\author{Jolien D. E. Creighton}
	\affiliation{Leonard E.\ Parker Center for Gravitation, Cosmology, and Astrophysics, University of Wisconsin-Milwaukee, Milwaukee, WI 53201, USA}

\author{Becca Ewing} 
	\affiliation{Department of Physics, The Pennsylvania State University, University Park, PA 16802, USA}
        \affiliation{Institute for Gravitation and the Cosmos, The Pennsylvania State University, University Park, PA 16802, USA}

\author{Heather Fong} 
	\affiliation{The University of Tokyo, Hongo 7-3-1 Bunkyo-ku, Tokyo 113-0033, Japan}

\author{Patrick Godwin} 
	\affiliation{Department of Physics, The Pennsylvania State University, University Park, PA 16802, USA}
        \affiliation{Institute for Gravitation and the Cosmos, The Pennsylvania State University, University Park, PA 16802, USA}

\author{Rachael Huxford}
	\affiliation{Department of Physics, The Pennsylvania State University, University Park, PA 16802, USA}
        \affiliation{Institute for Gravitation and the Cosmos, The Pennsylvania State University, University Park, PA 16802, USA}

\author{Shasvath Kapadia} 
	\affiliation{International Centre for Theoretical Sciences, Tata Institute of Fundamental Research, Bengaluru 560089, India}

\author{Alvin K. Y. Li}
	\affiliation{LIGO Laboratory, California Institute of Technology, MS 100-36, Pasadena, California 91125, USA}

\author{Rico Ka Lok Lo}
	\affiliation{LIGO Laboratory, California Institute of Technology, MS 100-36, Pasadena, California 91125, USA}

\author{Duncan Meacher} 
	\affiliation{Leonard E. Parker Center for Gravitation, Cosmology, and Astrophysics, University of Wisconsin-Milwaukee, Milwaukee, WI 53201, USA}

\author{Cody Messick} 
	\affiliation{Department of Physics, University of Texas, Austin, TX 78712, USA}
	
\author{Siddharth R. Mohite} 
	\affiliation{Leonard E. Parker Center for Gravitation, Cosmology, and Astrophysics, University of Wisconsin-Milwaukee, Milwaukee, WI 53201, USA}

\author{Atsushi Nishizawa}
	\affiliation{The University of Tokyo, Hongo 7-3-1 Bunkyo-ku, Tokyo 113-0033, Japan}

\author{Hiroaki Ohta}
	\affiliation{The University of Tokyo, Hongo 7-3-1 Bunkyo-ku, Tokyo 113-0033, Japan}

\author{Alexander Pace}
	\affiliation{Department of Physics, The Pennsylvania State University, University Park, PA 16802, USA}
        \affiliation{Institute for Gravitation and the Cosmos, The Pennsylvania State University, University Park, PA 16802, USA}

\author{Amit Reza}
	\affiliation{Department of Physics, Indian Institute of Technology Gandhinagar, Gujarat 382355, India}

\author{B.S. Sathyaprakash}
	\affiliation{Department of Physics, The Pennsylvania State University, University Park, PA 16802, USA}
        \affiliation{Institute for Gravitation and the Cosmos, The Pennsylvania State University, University Park, PA 16802, USA}
	\affiliation{Department of Astronomy \& Astrophysics, Pennsylvania State University, University Park, PA, 16802, USA}
	\affiliation{School of Physics and Astronomy, Cardiff University, Cardiff, UK, CF24 3AA}

\author{Minori Shikauchi} 
	\affiliation{The University of Tokyo, Hongo 7-3-1 Bunkyo-ku, Tokyo 113-0033, Japan}

\author{Divya Singh}
	\affiliation{Department of Physics, The Pennsylvania State University, University Park, PA 16802, USA}
        \affiliation{Institute for Gravitation and the Cosmos, The Pennsylvania State University, University Park, PA 16802, USA}

\author{Leo Tsukada}
	\affiliation{The University of Tokyo, Hongo 7-3-1 Bunkyo-ku, Tokyo 113-0033, Japan}

\author{Daichi Tsuna}
	\affiliation{The University of Tokyo, Hongo 7-3-1 Bunkyo-ku, Tokyo 113-0033, Japan}

\author{Takuya Tsutsui}
	\affiliation{The University of Tokyo, Hongo 7-3-1 Bunkyo-ku, Tokyo 113-0033, Japan}

\author{Koh Ueno}
	\affiliation{The University of Tokyo, Hongo 7-3-1 Bunkyo-ku, Tokyo 113-0033, Japan}

\begin{abstract}
Binary neutron stars (BNSs) will spend $\simeq 10$ -- 15 minutes in the band of
Advanced LIGO and Virgo detectors at design sensitivity. Matched-filtering of
gravitational-wave (GW) data could in principle accumulate enough
signal-to-noise ratio (SNR) to identify a forthcoming event tens of seconds
before the companions collide and merge. Here we report on the design and
testing of an early warning gravitational-wave detection pipeline. Early
warning alerts can be produced for sources that are at low enough redshift so
that a large enough SNR accumulates $\sim 10 - 60\,\rm s$ before merger. We
find that about 7\% (respectively, 49\%) of the total detectable BNS mergers
will be detected $60\, \rm s$ ($10\, \rm s$) before the merger. About 2\% of
the total detectable BNS mergers will be detected before merger and localized
to within $100\, \rm \text{deg}^2$ (90\% credible interval).  Coordinated
observing by several wide-field telescopes could capture the event seconds
before or after the merger. LIGO-Virgo detectors at design sensitivity could
facilitate observing at least one event at the onset of merger.
\end{abstract}

\section{Introduction} \label{sec:intro}
August 17, 2017 saw the beginning of a new era in multi-messenger astronomy
with the joint detection of GWs by the LIGO and Virgo
interferometers and the gamma-ray burst by the Fermi-GBM and INTEGRAL satellite
from the BNS coalescence, GW170817~\citep{TheLIGOScientific:2017qsa}. The
detection was followed by observations of the electromagnetic (EM) counterpart
and afterglow by gamma-ray, UV, optical, infra-red, and radio telescopes. These
observations triggered several important science results: (a) they settled a
long-standing question about the origin of short gamma-ray
bursts~\citep{GBM:2017lvd}, (b) provided a new tool for measuring cosmological
parameters (with the first measurement of the Hubble constant using standard
sirens~\citep{Abbott:2017xzu}), (c) confirmed the production of heavy elements
in the aftermath of the merger~\citep{Abbott:2017wuw}, (d) triggered many
questions about the central engine producing gamma-ray burst (GRB) and
afterglows~\citep{Monitor:2017mdv}, and (e) set limits on the difference in the
speed of GWs and light helping rule out certain alternative
theories of gravity~\citep{Monitor:2017mdv, LIGOScientific:2019fpa}. 

Apart from the gamma-ray burst, which was observed $\sim 2$~s after the merger
event, the first manual follow-up observations took place $\sim 8$ hours after
the epoch of merger~\citep{GBM:2017lvd}. The GW alert was sent out $\sim 40$
minutes ~\citep{GCN21505} and the sky localization $\sim 4.5$
hours~\citep{GCN21513} after the signal arrived on earth. Among the factors
that contributed to the delay were a non-stationary glitch in the Livingston
interferometer and issues with the transfer of data from the Virgo detector to
analysis sites delaying the sky localization of the event.  By the time EM
telescopes participating in the follow-up program received the alerts the
source was below the horizon for them. 

For a fraction of BNS events it will be possible to issue
alerts up to $\delta t \sim 60\,\rm s$ before the epoch of merger.
Pre-merger or \emph{early warning} detections will facilitate electromagnetic
observations of the prompt emission, which encodes the initial conditions of
the outflow and the state of the merger remnant.  Indeed, early optical and
ultraviolet observations are necessary to further inform our understanding of
\emph{r}-process nucleosynthesis~\citep{Nicholl:2017ahq} and shock-heated
ejecta~\citep{Metzger:2017wot}, while prompt X-ray emission would reveal the
final state of the
remnant~\citep{Metzger:2013cha,Ciolfi:2014yla,Siegel:2015twa}.  Early
observations made in the radio band could indicate pre-merger magnetosphere
interactions~\citep{Most:2020ami}, and would test models that propose BNS as a
possible precursor of fast radio
bursts~\citep{Totani:2013lia,Wang:2016dgs,Dokuchaev:2017pkt}.

The \texttt{GstLAL}-based inspiral
pipeline~\citep{Sachdev:2019vvd,Hanna:2019ezx,Messick:2016aqy} (hereafter
shortened to \texttt{GstLAL}) is a low-latency matched-filtering pipeline used
to detect gravitational waves from compact binary coalescences in LIGO-Virgo
data. Other low-latency detection pipelines running on LIGO-Virgo data include
\texttt{PyCBCLive}~\citep{Nitz:2018rgo},
\texttt{MBTAOnline}~\citep{Adams:2015ulm}, SPIIR~\citep{chu2017low}, and the
unmodeled search -- \texttt{CWB}~\citep{Klimenko:2015ypf}. In a seminal
paper,~\citet{Cannon:2011vi} described a computationally practical filtering
strategy for near real time matched-filtering of GW data that could produce
early-warning triggers. This work describes the foundations of
\texttt{GstLAL}, which has been detecting GWs in low-latency since the first
observing run (O1) of the Advanced LIGO and Virgo detectors.
\citet{Cannon:2011vi} also discussed the expected rates of BNS events that
could be detectable before merger and prospects for their localizations based
on theoretical SNR and Fisher estimates. There have been other studies
examining theoretical potentials of pre-merger BNS detections, such
as~\citet{10.1093/mnras/stw576}. In this \FIXME{Letter}, for the first time, we
show the implementation of a search which can detect BNSs pre-merger and
provide early warnings to other observatories in practice; we examine the
performance of \texttt{GstLAL} in recovering BNS systems before merger by
running it over a month of simulated data with added signals. Based on the
median rate of BNS mergers deduced from GW170817 and
GW190425~\citep{Abbott:2020uma}, our studies suggest alerts could be issued
\FIXME{10 s (60 s)} before merger for \FIXME{24 (3)} BNS systems over the
course of one year of observations of a three-detector Advanced network
operating at design sensitivity. Our results broadly agree with the estimates
of ~\citet{Cannon:2011vi}. In addition, we provide the distribution of
realistic sky localizations (all sky localizations quoted are 90\% credible
intervals unless stated otherwise) for various times before merger, using a
rapid Bayesian localization tool, \texttt{BAYESTAR}~\citep{PhysRevD.93.024013}.
We find that based on current BNS merger rate estimates, $\mathcal{O}(1)$ event
will be both detected before merger and localized to $100 \, \text{deg}^2$.

These results assume zero latency from data transfer, calibration, filtering,
and follow-up processes. In real application, these latencies will need to be
subtracted from the pre-merger times at which we can provide alerts. In the
latest observing run of the Advanced LIGO-Virgo detectors (O3), these latencies
accounted for $\sim 20\, \rm s$ of delay in alerts, but eventutally we hope to
be able to reduce it to $\sim 7\, \rm s$ for the early warning alerts. The
remainder of the paper is structured as follows: we discuss the pipeline and
simulations used in Section~\ref{sec:simulation}, the prospects of rapid sky
localization of pre-merger candidates in Section~\ref{sec:localization}, and
broader implications of coincident GW and EM observation in
Section~\ref{sec:discussion}.

\section{Simulation} \label{sec:simulation}
We assess the prospects of pre-merger alerts with an Advanced LIGO-Virgo
network at design sensitivity. For this study, we generate one month of
stationary Gaussian data recolored to Advanced LIGO and Advanced Virgo design
sensitivities\footnote{We use the power spectral densities provided in
https://dcc.ligo.org/LIGO-T0900288/public and
https://dcc.ligo.org/LIGO-P1200087/public for the Advanced LIGO and Advanced
Virgo interferometers, respectively. We assume that LIGO-Hanford and
LIGO-Livingston will reach the same design sensitivities.}.

We generate a population of \TotalSims{} simulated BNS signals, henceforth
referred to as \textit{injections}, using the \texttt{SpinTaylorT4} waveform
model~\citep{Buonanno:2004yd}. Both source-frame component masses are drawn
from a Gaussian distribution between \SimMinMass{} $< m_1, m_2 <$ \SimMaxMass{}
with mean mass of \SimMeanMass{} and standard deviation of \SimStdev{}, modeled
after observations of galactic BNSs~\citep{Ozel:2016oaf}.  The neutron stars in
the population are non-spinning, motivated by the low spins of BNSs expected to
merge within a Hubble time~\citep{Burgay:2003jj,Zhu:2017znf}. The signals are
distributed uniformly in comoving volume up to a redshift of $z =
\SimMaxRedshift{}$. We reject \NumRejectedSims{} injections with LIGO-Hanford
or LIGO-Livingston SNRs below 3 to reduce the computational load and inject the
remaining signals in the Gaussian data. We don't expect the search to recover
signals with such small SNRs so no bias is introduced in rejecting these.

We use the offline configuration of the \texttt{GstLAL}
pipeline~\citep{Sachdev:2019vvd,Hanna:2019ezx,Messick:2016aqy} to recover the
remaining \NumAcceptedSims{} injected into the Gaussian data described above. 
\texttt{GstLAL} has been successfully detecting compact binary coalescences in low-latency since O1
~\citep{Abbott:2016nmj} and is so far the only pipeline to detect a BNS in
low-latency~\citep{TheLIGOScientific:2017qsa, Abbott:2020uma}.

\subsection{\texttt{GstLAL} methods} \label{ssec:methods}
Matched-filtering GW searches use a template
bank~\citep{Owen:1998dk} containing a set of GW  waveforms covering the desired
parameter space. \texttt{GstLAL} divides the template bank into several sub-banks by
grouping templates that respond to noise in a similar fashion based on their
intrinsic parameters~\citep{Messick:2016aqy,Sachdev:2019vvd}. It then uses the
LLOID method~\citep{Cannon:2011vi} to construct orthogonal basis filters from
the sub banks by performing in order multi-banding and singular value
decomposition (SVD)~\citep{Cannon_2010} of each time slice. The data is
cross-correlated with the basis filters to produce GW
candidates.
Candidates with SNRs below \SNRThresh{} are discarded
to reduce the volume of triggers.  Candidates that survive this step are
assigned a log likelihood-ratio, \likelihood. The log likelihood-ratio ranks
candidate events by their SNR, the sensitivity of each detector at the time of
the trigger, an autocorrelation-based signal consistency test ($\xi^2$), and (for
coincident triggers) the time and phase delays between participating
interferometers~\citep{Cannon:2015gha,Messick:2016aqy,Hanna:2019ezx,Sachdev:2019vvd}.
A template-dependent factor, $\log P(\vec{\theta}_k|\text{signal})$ where
$\vec{\theta}$ denotes the template, is included in the log likelihood-ratio to
account for the population mass-model~\citep{Fong:2018elx} of signals. 
The distribution of log-likelihood ratio for noise triggers is created by
sampling the noise distributions of the parameters it depends on, and all
candidates are subsequently assigned a false-alarm-rate to describe how often a
candidate with a \likelihood{} at least as high as its own is expected to be
produced from noise fluctuations.

\subsection{Early-warning methods} \label{ssec:ewmethods}
In this search, we used a stochastically generated template
bank~\citep{Privitera:2013xza,Harry:2009ea} with non-spinning components between 
masses \BankMinMass{} $<m_1, m_2<$ \BankMaxMass{}; bounds chosen to account for
edge effects and redshift, and chirp mass $\in$ (\BankMinChirp{},
\BankMaxChirp{}); bounds chosen based on the Gaussian population described
above. We model the GW emission from 10 Hz to merger using the
\texttt{TaylorF2}~\citep{Sathyaprakash:1991mt, Blanchet:1995ez,
Blanchet:2005tk, Buonanno:2009zt} waveform. The resulting template bank has a
minimum match of \BankMM{}~\citep{Owen:1998dk} and consists of \BankSize{} waveforms. The
template-dependent factor to account for the population mass-model used in the
log-likelihood ratio is modeled as a Gaussian in chirp mass with a mean of
$1.18 M_\odot$ and a standard deviation of $0.055 M_\odot$. The mean chirp mass
is derived from the Gaussian component mass distribution described in Section~\ref{sec:simulation} 
at a redshift of $z = 0.02$.
  
We repeat the search six times, using the same template bank and the same
dataset inluding injections, to determine the pipeline's performance at various
times before merger. The searches begin filtering at 10 Hz, but complete
filtering at different frequencies. In particular, we choose $29\, \rm Hz$,
$32\, \rm Hz$, $38\, \rm Hz$, $49\, \rm Hz$, $56\, \rm Hz$, and $1024\, \rm Hz$
to analyze signal recovery at (approximately) $58\, \rm s$, $44\, \rm s$, $28\,
\rm s$, $14\, \rm s$, $10\, \rm s$, and $0\, \rm s$ before merger. We will
refer to each ending frequency configurations as a different ``run" in the
discussion that follows. In practice, these ending frequencies are only
approximate, since we chose to align the waveforms that are grouped together
before performing the SVD (see Section.~\ref{ssec:methods}) such that the
waveforms in each sub bank provide the same pre-merger time. The times before
merger quoted here are the median times for each run.  In our simulation, this
time ranged from \FIXME{$\sim 6 \, \rm s - 99 \, \rm s$} between the 6 runs.
While performing multi-banding, the waveforms belonging to a sub-bank are
time-sliced and each slice is sampled according to the highest Nyquist
frequency in that sub-bank and time slice. However, for these runs we fixed the
sample rate of the final time-slice at \FIXME{$2048 \, \rm Hz$} so that the
$\xi^2$ is calculated at the full frequency resolution. As the bandwidth of the
search is decreased, the variance associated with the recovered end time, phase,
and SNR grow. We account for increased uncertainty in the signal end time by
extending the time window in which we search for coincident signals to 10
milliseconds plus light travel time. We repeat the procedure described
in~\cite{Hanna:2019ezx} for each analysis to account for bandwidth related
changes to the covariance matrix and construct signal distributions for time
and phase delays for each of the run. In addition, we tuned the binning and
sampling of the SNR and $\xi^2$ histograms which are used to calculate the
distribution of log-likelihood ratio of noise triggers which defines the background
model of the search. In absence of any simulated signals in the
Gaussian data, we expect the foreground of our runs to agree with the background model
computed by the search. We confirmed that for each of the 6 runs, on excluding
the simulated signals, the distribution of log-likelihood ratio of the
candidates agreed with the background model computed by the search.

\subsection{Results} \label{ssec:results}
We consider any injection that is recovered with a FAR $<=1/(30\, \rm days)$ to
be \textit{found} by our pipeline in each of the 6 different runs.  We can compute the expected number of signals for each run
based on the sensitive spacetime volume of each run at our chosen FAR threshold
and on the local BNS merger rate, \FIXME{$250-2810\, \text{Gpc}^{-3} \,
\text{a}^{-1}$} (90\% credible interval) \citep{Abbott:2020uma}.  The
sensitive spacetime volume of the search at a given FAR threshold is then
estimated as
\begin{align}
\langle VT \rangle = \langle VT\rangle_{\mathrm{injected}}
\frac{N_{\mathrm{recovered}}}{N_{\mathrm{total \, sims}}},
\end{align}
where $N_{\mathrm{recovered}}$ is the number of recovered injections at the
given FAR. This assumes that the injections have not been restricted to space
or time that the pipeline could gave been sensitive to. For this signal
distribution, the simulated signals probed a spacetime volume of
$VT_{\mathrm{injected}} = \FIXME{0.178\,\mathrm{Gpc}^3\,\mathrm{a}}$. The
results are shown in Table~\ref{table:results}.  We expect \FIXME{12--132} BNSs
per year for a three-detector Advanced network at design sensitivity, about
half of which will be detected \FIXME{$10\, \rm s$} before merger and
\FIXME{1--9} events will be detected a minute before merger.

At a FAR of \FIXME{1/30 days}, based on the current median BNS merger rate
(\FIXME{$1035\, \text{Gpc}^{-3} \, \text{a}^{-1}$}, average of the two median
rates from~\citet{Abbott:2020uma}), the contamination fraction from noise for
the $29\, \rm Hz$ run ($60\, \rm s$) before merger is \FIXME{79\%}, going down
to \FIXME{20\%} for the full bandwidth run. The contamination fraction is
higher for runs that provide the earliest triggers.  This suggests a natural
method to vet early warning triggers; triggers that are identified in an
`early' band but not later bands are likely to be noise.

\squeezetable
\begin{table}[t]
\begin{tabular}[t]{cccc}
\toprule

$f_{\rm high}$ (\rm Hz) & $\langle VT \rangle (\mathrm{Gpc}^3\,\mathrm{a})$ & $N_{\rm signals} (\rm {a}^{-1})$ &  $N_{\rm low} - N_{\rm high}(\rm {a}^{-1})$ \\
\midrule
29 & $2.55 \times 10^{-4}$ & 3.21 & 0.775 -- 8.71\\
32 & $3.84 \times 10^{-4}$ & 4.84 & 1.17 -- 13.2\\
38 & $7.23 \times 10^{-4}$ & 9.12 & 2.20 -- 24.8\\
49 & $1.45 \times 10^{-3}$ & 18.2 & 4.41 -- 49.5\\
56 & $1.88 \times 10^{-3}$ & 23.6 & 5.71 -- 64.2\\
1024 & $ 3.86 \times 10^{-3}$ & 48.7 & 11.8 -- 132\\
\bottomrule
\end{tabular}
\caption{\label{table:results}\FIXME{Sensitive spacetime volume ($\langle VT
\rangle$) of the 6 runs and the expected number of signals
($N_{\rm signals}$) per year based on the median BNS merger rate. We also show
the expected range of events based on the uncertainty in the BNS merger rate
$(N_{\rm low} - N_{\rm high})$.}}
\end{table}

\section{Sky Localization of Early Warning Alerts} \label{sec:localization} 
\begin{figure}
\includegraphics[width=1.0\linewidth]{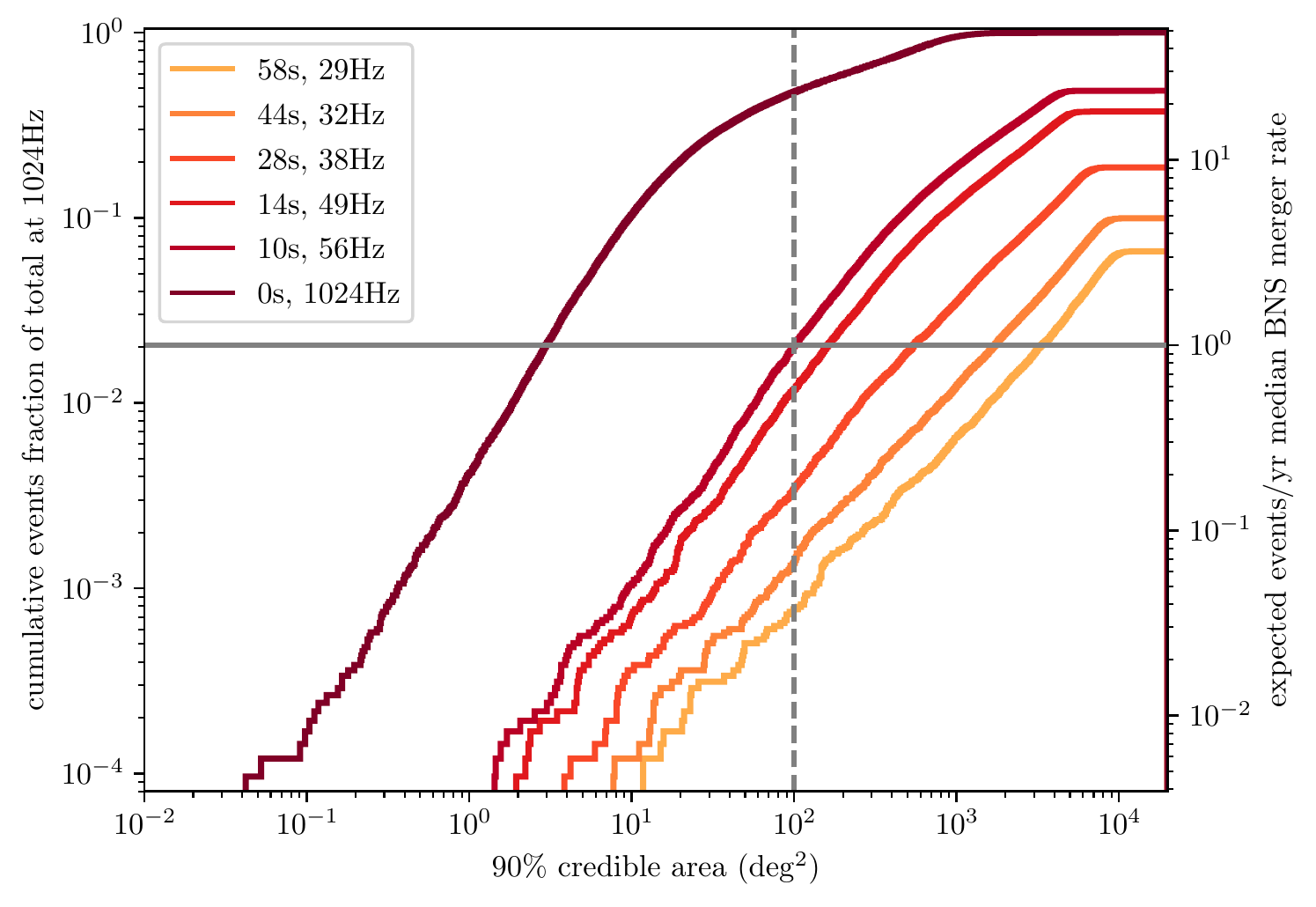}
\caption{\label{fig:area_hist}\FIXME{Cumulative distributions of the sky
localizations (90\% credible interval) of injections that pass the FAR
threshold in each run. Results are shown for a three-detector
network (LIGO-Hanford, LIGO-Livingston, Virgo) operating at design sensitivity.
The left vertical axis shows the number as a fraction of the total recovered
injections at full bandwidth. The right vertical axis shows the expected number of
events per year based on the median BNS merger rate. We expect at least one
event per year detected before merger and localized to within $100\,
\text{deg}^2.$}}
\end{figure}

\begin{figure}
\includegraphics[width=1.0\linewidth]{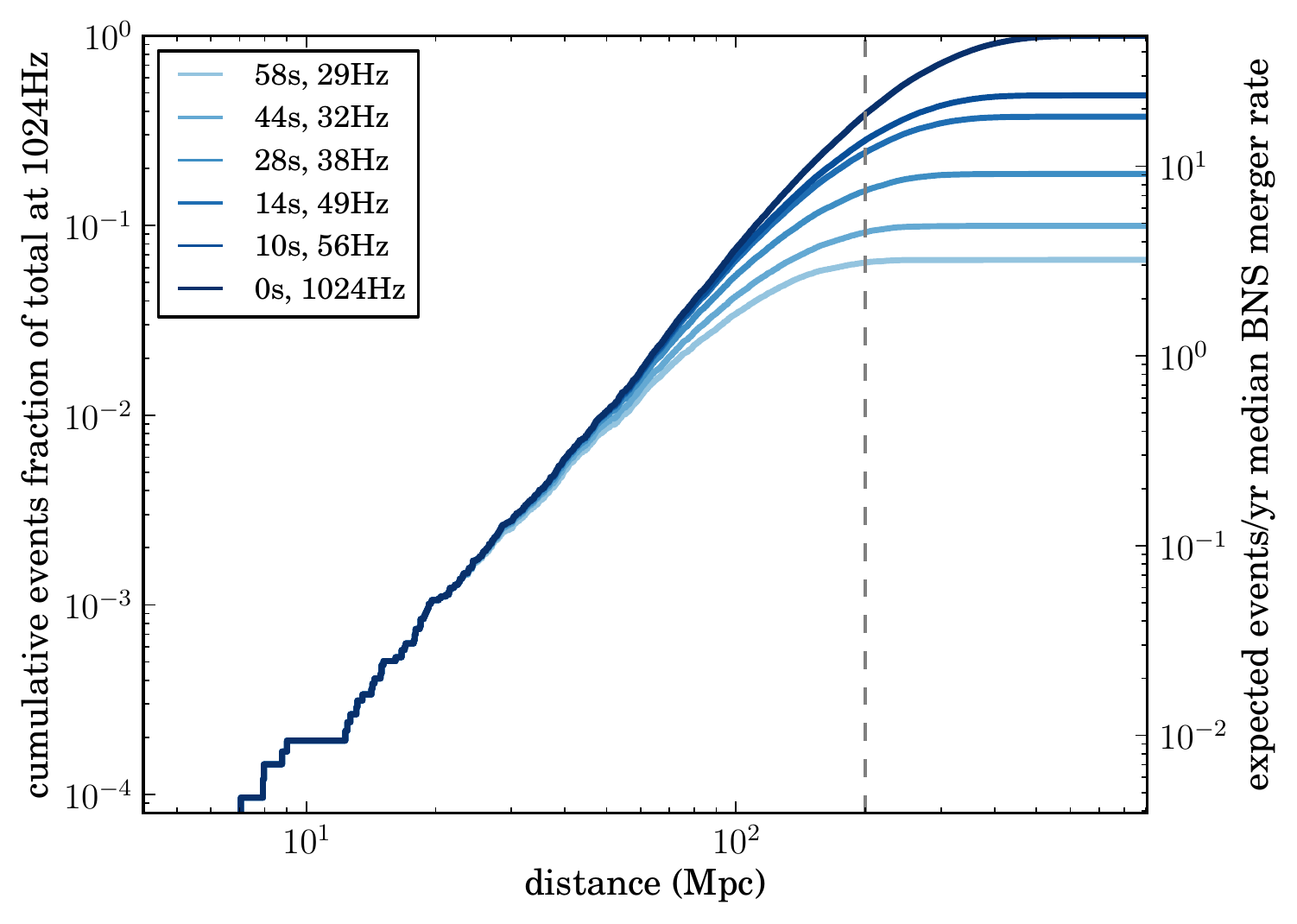}
\caption{\label{fig:dist_hist}\FIXME{Cumulative distributions of the luminosity
distance of injections that pass the FAR threshold in each run.
Results are shown for a three-detector network (LIGO-Hanford, LIGO-Livingston,
Virgo) operating at design sensitivity. The left vertical axis shows the number as a
fraction of the total recovered injections at full bandwidth. The right vertical axis
shows the expected number of events per year based on the median BNS merger
rate. About half of the events that are detected before merger will be
within $200 \, \rm Mpc$.}}
\end{figure}

The
primary goal of providing pre-merger alerts for BNSs is to facilitate
electromagnetic observations before and/ or at merger. The earliest alerts we
could provide will be $\sim 60\, \rm s$ before merger, therefore to achieve
this goal it is crucial that we provide rapid and accurate sky localizations.
LIGO-Virgo use \texttt{BAYESTAR}~\citep{PhysRevD.93.024013} to generate rapid
localizations, which is a fast Bayesian algorithm that can reconstruct
positions of GW transients using the output provided by the matched-filtering
searches. We generate the SNR time series of all injections that pass the FAR
threshold for each run and provide these to \texttt{BAYESTAR} in order to localize the
signals. The results are shown in Fig.~\ref{fig:area_hist}.  We show the
cumulative histograms of the \FIXME{90\%} credible interval of sky
localizations of the injections that pass the FAR threshold in each run. The
right vertical axis shows the expected number of events per year as a function
of the largest localization area based on the median merger rate, the left
vertical axis shows this number as a fraction of the total injections recovered
at the full bandwidth. In addition we show cumulative histograms of luminosity
distances of events that pass the FAR threshold in Fig.~\ref{fig:dist_hist}.
These results can be easily reinterprated with any update in the BNS merger
rate.

\section{Discussion} \label{sec:discussion}
Ideally we want the signals to be well localized in sky given the small fields
of view (FOVs) of optical telescopes. Fig.~\ref{fig:area_hist}
shows that at least one event per year will be both detected before merger and
localized to within \FIXME{$100 \, \text{deg}^2$}. Furthermore, if we consider
the ``searched area", defined as the area searched in the sky according to the
localization PDF before finding the true location of the event, about \FIXME{9}
events per year ($\sim 18\,\%$ of total detectable BNSs) will be both detected
before merger and found before searching over $100 \, \text{deg}^2$.
Additionally, the searched area can be reduced by using galaxy catalogs to
inform imaging strategies~\citep{Hanna_2014}. Events we are able to provide early warnings for, especially the well localized
ones, will be the ones that are the closest to us further enabling better
follow-up. At least \FIXME{1 event per year} (3.4\% of the total) detected
$60\, \rm s$ before merger will be within $100\, \rm Mpc$ and about \FIXME{13
events per year} ($28 \%$ of the total) will both be detected before merger and
lie within $200 \, \rm Mpc$ (Fig.~\ref{fig:dist_hist}).  

Wide-field optical transient facilities such as the BlackGEM array ($0.65\, \rm
m /2.7 \, \text{deg}^2$ per telescope) with 3 telescopes planned in the first
phase of operation eventually exapnding to 15 telescopes~\citep{blackgem}, the
Zwicky Transient Facility ($1.2 \, \rm m / 47 \, \text{deg}^2$)~\citep{ztf},
the Dark Energy Camera ($4\, \rm m / 3.8 \,
\text{deg}^2$)~\citep{Flaugher_2015}, the Rubin Observatory ($8.4 \, \rm m /
9.6 \, \text{deg}^2$)~\citep{lsst}, the Swope Telescope ($1\, \rm m / 7 \,
\text{deg}^2$)~\citep{swope}, the Subaru Telescope ($8.2 \, \rm m/1.7 \,
\text{deg}^2$)~\cite{subaru}, etc., operated in ``target of opportunity" mode
will be most fitting for the optical follow-up of well localized events. Events
with larger localization areas will be useful to alert space telescopes, such
as the Fermi Gamma-ray Burst Monitor (all-sky) and the Swift Observatory
(hosting the Burst Alert Telescope with a FOV of $\sim 10000 \, \text{deg}^2$
and can localize events with an accuracy of 1 to 4 arc-minutes within $15\, \rm
s$, X-Ray Telescope, and Ultra Violet Optical Telescope). Pre-merger GW detection will be
especially helpful to identify subthreshold GRBs for off-axis BNS mergers~\citep{Tohuvavohu_2020}. Radio telescopes with large FOVs of hundreds of square degrees such as the
Murchison Widefield Array~\citep{mwa}, the National Radio Astronomy
Observatory~\citep{nrao} consisting of several telescope arrays, the Giant
Metre-Wave Radio Telescope~\citep{gmrt}, the Owens Valley Long Wavelength
Array~\citep{ov-lwa}, and (under construction) the Square Kilometre
Array~\citep{ska}, etc. can use even the poorly localized early warning alerts. ~\citet{Callister_2019} have demonstrated a search which
looks for radio signals coincident with GW alerts by buffering the data of
OVRO-LWA, to look for signals coincident with the
GW170104~\citep{Abbott:2017vtc}. Early warning alerts will enable buffering of
the radio data at significantly higher time
resolution.~\citet{10.1093/mnrasl/slz129} also describe using negative-latency
BNS merger alerts to detect prompt radio bursts with MWA. Early warning alerts
will also be useful for ground-based gamma-ray detector facilities such as the
Cherenkov Telescope Array, which consists of several fast slewing
telescopes~\citep{cta}. They can slew within tens of seconds but must be
pointing at the source at the time of merger, since there is no hypothesized
after glow at the high energies these telescopes can detect.   

The pre-merger latencies described in the paper are
the median latencies from each run. The exact pre-merger latencies
depend on the template masses, and range from \FIXME{$6 \, \rm s$--$99 \, \rm
s$}.  The pre-merger times quoted in this paper also assume zero latency from
data transfer, calibration, and the matched-filtering processes. In O3, this
latency was about $\sim 20 \, \rm s$; our goal is to bring this latency down to
$\sim 7 \, \rm s$ for the smaller bandwidth (early warning) configurations.

The low-latency \texttt{GstLAL} pipeline recently participated in a test of the
LIGO-Virgo early warning infrastructure and issued the first test alerts and
retractions for pre-merger candidates~\citep{GCN27951, user-guide}. This test will be
described in more detail in a future publication.

\section{Acknowledgements}
This work was supported by the NSF grant OAC-1841480.  S.S. is supported by the
Eberly Research Funds of Penn State, The Pennsylvania State University,
University Park, Pennsylvania. D.M. acknowledges the support of NSF
PHY-1454389, ACI-1642391, and OAC-1841480. S.R.M thanks the LSSTC Data Science
Fellowship Program, which is funded by LSSTC, NSF Cybertraining Grant
\#1829740, the Brinson Foundation, and the Moore Foundation; his participation
in the program has benefited this work. B.S.S. is supported in part by NSF
Grant No. PHY-1836779 and AST-1708146. Computations for this research were
performed on the Pennsylvania State University’s Institute for Computational
and Data Sciences Advanced CyberInfrastructure (ICDS-ACI). We also thank the
LIGO Laboratory for use of its computing facility to make this work possible.

\software{The analysis of the data and the detections of the simulation signals
were made using the \textsc{GstLAL}-based inspiral software pipeline
\citep{Cannon:2011vi, Privitera:2013xza, Messick:2016aqy, Sachdev:2019vvd,
Hanna:2019ezx}. These are built on the \textsc{LALSuite} software library
\citep{LALSuite}. The sky localizations made use of
\texttt{ligo.skymap}\footnote{https://lscsoft.docs.ligo.org/ligo.skymap}, which
uses Astropy,\footnote{http://www.astropy.org} a community-developed core
Python package for Astronomy \citep{astropy:2013, astropy:2018}. The plots
were prepared using Matplotlib \citep{2007CSE.....9...90H}.}

\bibliography{references}

\end{document}